\relax
\documentstyle [12pt]{article}
\newcommand{\la}{\langle}
\newcommand{\ra}{\rangle}

\newcommand{\p}{\partial}
\begin {document}
\bibliographystyle {plain}

\title{\bf One-dimensional spin-liquid without magnon excitations}
\author{A.A.Nersesyan$^1$ and A.M.Tsvelik$^2$}
\maketitle
\begin {verse}
$^1{\em International ~Centre ~for ~Theoretical ~Physics,~P.O.Box~586,~34100,~Trieste,}$\\ 
${\em Italy,~ and}$\\
${\em Institute ~of ~Physics, ~Tamarashvili ~6, ~380077, ~Tbilisi, ~Georgia}$\\
$^2{\em Department~ of~ Physics,~ University~ of~ Oxford,~ 1~ Keble~ Road,}$\\
${\em~Oxford,~OX1~ 3NP,~ UK}$
\end{verse}
\begin{abstract}
\par

It is shown that a sufficiently strong four-spin interaction 
 in the spin-1/2 spin
ladder can cause dimerization. Such interaction can
be generated  either by  phonons or (in the doped state) by the
conventional Coulomb repulsion between the holes. 
The dimerized phases are  thermodynamically undistinguishable from the Haldane
phase, but  have  dramatically  different correlation functions: the
dynamical magnetic susceptibility, instead of  displaying a 
sharp single magnon peak near $q = \pi$, shows only a two-particle threshold
separated from the ground state by a gap. 

\end{abstract}

\sloppy


The qualitative difference between the universal properties of one-dimensional
Heisenberg antiferromagnets with half-integer
and integer spin, originally predicted by Haldane \cite{Haldane}, 
is now well understood.
Typical examples are the S = 1/2 and S = 1 chains.
The spin S = 1/2 chain has a quasi-ordered singlet ground state with algebraically decaying spin 
correlations. Its spectrum is gapless, and the elementary excitations (spinons), carrying spin 1/2, 
in all physical states with integer total spin appear only in
pairs \cite{Faddeev}. Deconfinement of the spinons implies 
that conventional S = 1 magnons fail to be
stable quasiparticles: the spectral density of the staggered magnetization
${\bf n} (x) \sim (-1)^n {\bf S}_n $ shows a purely incoherent background.

An alternative picture of a disordered spin liquid with a Haldane gap 
is effectively realized
when two S = 1/2 Heisenberg chains are put together to form a spin
ladder 
\cite{Dagotto}. 
In the standard model of spin ladders the interchain interaction is the
Heisenberg exchange $J_{\perp}$; at $J_{\perp} \neq 0$ 
the spinons confine to form triplet (magnon) and singlet
excitations  with gaps $m_t$ and $m_s$ ($m_t , m_s \sim J_{\perp}$). 
The triplet excitations  contribute a coherent 
$\delta$-peak to
the dynamical spin susceptibility $\chi''(q, \omega)$ near  $q = \pi$ and
$\omega = m_t$ \cite{Shelton1}. In this respect  the spin liquid behaves
like a conventional magnet with the only difference that here the
magnons are ``optical'', that is have a spectral gap.

In this letter  we discuss  an example of a disordered spin liquid 
with a gapful spectrum but {\it without} coherent magnon excitations. 
Such a state can be realized in an extended
model of the spin S = 1/2 Heisenberg ladder which, apart from 
the direct transverse exchange
$J_{\perp}$, also includes a four-spin interaction
\begin{eqnarray}
{\cal H} &=& J_{\parallel} \sum_{j=1,2} \sum_n {\bf S}_j (n) \cdot {\bf S}_j (n + 1) \nonumber\\
&+& \sum_n [ J_{\perp} {\bf S}_1 (n) \cdot {\bf S}_2 (n) + 
V ({\bf S}_1 (n) \cdot {\bf S}_1 (n + 1)) ({\bf S}_2 (n) \cdot {\bf S}_2 (n + 1)) ].
\label{gen.model}
\end{eqnarray}
It will be assumed that $|J_{\perp}|, |V| \ll J_{\parallel}$.
The new term in (\ref{gen.model}) represents an interchain coupling of the spin-dimerization fields
$\epsilon_j (n) = (-1)^n {\bf S}_j (n) \cdot {\bf S}_j (n + 1)~(j =
1,2)$ which  can be either (i)
effectively mediated by spin-phonon interaction (see below) or (ii) in the doped phase  
generated by the conventional Coulomb repulsion
between the holes moving in the spin correlated background (\cite{Shelton2}). 

 In a single S = 1/2 Heisenberg chain described in the continuum limit in terms of
a massless scalar field $\Phi_s (x)$ \cite{Affleck}, the dimerization operator
$
\epsilon (x) \sim \cos \sqrt{2 \pi} \Phi_s (x)
$
is a strongly fluctuating field with the same scaling dimension 1/2 as that of the three components
of the staggered magnetization
$
(-1)^n {\bf S}_n \sim {\bf n} (x) \sim (\cos \sqrt{2 \pi} \Theta_s (x), 
\sin \sqrt{2 \pi} \Theta_s (x), - \sin \sqrt{2 \pi} \Phi_s (x) ),
$
where $\Theta_s$ is the field dual to $\Phi_s$. Therefore the interchain coupling term 
$V \epsilon_1 \epsilon_2$, being as relevant as the transverse exchange
$J_{\perp} {\bf n}_1 \cdot {\bf n}_2$, may significantly affect the low-energy properties of spin
ladders. Indeed, while for small $|V|$ the generalized model
(\ref{gen.model}) occurs in the same Haldane spin liquid phase as that of
the conventional spin ladder ($V = 0$), at large enough $|V|$  
the model (\ref{gen.model}) displays {\it non-Haldane} spin liquid
phases characterized by a
{\it spontaneous} dimerization. The thermodynamic properties of these phases,
determined by the absolute values of the gaps,
are indistiguishable from those for the Haldane phase. However, the correlation
functions of the two systems differ drastically: in a non-Haldane spin
liquid we
are going to discuss the spectral function of ${\bf n}$ is entirely exhausted by an incoherent background.
This picture should be opposed to the dimerized (spin-Peierls) S = 1/2 Heisenberg chain with
alternating exchange $J_{n,n+1} = J [ 1 + \delta (-1)^n]$ where, due to the presence of the kink-antikink
bound states in the spin excitation spectrum, triplet 
magnons constitute well defined, coherent
quasiparticles.

It has been shown earlier \cite{Shelton1}, \cite{Shelton2} that, in the continuum limit, the Hamiltonian
(\ref{gen.model}) decouples into four massive real fermionic 
fields, or equivalently,
four noncritical 2D Ising models
\footnote{This decoupling is valid up to a weak residual (marginal) interaction whose role in the gapped
phases is
exhausted by the mass and velocity renormalization \cite{Shelton1}.},
exhibiting the underlying $SU(2) \times Z_2$ symmetry:
$
H = \sum_{a=1,2,3} H_{m_t} [\xi_a] + H_{m_s} [\xi_0],
$
where 
\begin{equation}
H_m [\xi] = - \frac{i v_s}{2} (\xi_R \p_x \xi_R - \xi_L \p_x \xi_L) - i m \xi_R \xi_L.
\end{equation}
The triplet and singlet masses are given as follows:
\begin{equation}
m_t = \pi a_0 (J_{\perp} |a_s|^2 - V |a_c|^2), ~~~
m_s = - \pi a_0 (3 J_{\perp} |a_s|^2 + V |a_c|^2).
\end{equation}
where $a_c$ and $a_s$ are nonuniversal constants. 
The thermodynamic properties are determined only by 
$|m_t|, |m_s|$, but the symmetry of the ground state and the behavior
of 
the dynamical sisceptibility $\chi (q, \omega)$ 
 crucially depend on the relative signs of the masses $m_t$ and $m_s$ as well.
This follows from the fact that the representation
for the staggered spin density and dimerization operators in terms of 
the Ising order and 
disorder variables\cite{Shelton1} depends on the sign of $m_t m_s$. 
It will be assumed below that $V < 0$, while the sign of $J_{\perp}$
can 
be arbitrary.
We shall show that for different signs of $m_t$ and $m_s$, as long as
the 
triplet branch of the spectrum
remains the lowest ($|m_t| < |m_s|$), the two-chain ladder is in the 
Haldane-liquid phase
with short-ranged correlations of the staggered magnetization and 
dimerization field, but
with coherent S = 1 and S = 0 single-magnon excitations. If $m_t$ and
$m_s$ 
have the same sign, the ground
state represents a dimerized, spin-disordered phase in which coherent 
magnon modes are absent.

Transitions from the Haldane to dimerized phases take place 
when either the triplet excitations
become gapless, with the singlet mode still having a finite gap, or vice versa. (Notice that
in the conventional spin ladder with $V = 0$, both masses vanish simultaneously at $J_{\perp} \rightarrow 0$,
and the ladder trivially decouples into a pair of independent S = 1/2 Heisenberg chains.)
The transition at $m_t = 0$ belongs to the universality class of the critical, exactly
integrable, S = 1 spin chain \cite{Babu}; the corresponding non-Haldane phase with
$|m_t| < |m_s|$ represents the dimerized state of the S = 1 chain with spontaneously broken translational
symmetry and doubly degenerate ground state \cite{Affleck1}. The critical point $m_s = 0$ is
of the Ising type; it is associated with a transition to another dimerized phase ($|m_t| > |m_s|$),
not related to the S = 1 chain.

We start our discussion by considering the standard $J_{\parallel}$-$J_{\perp}$ two-chain ladder,
with spin-phonon coupling included via a substitution
$
J_{\parallel} \rightarrow J_{\parallel, j} (n, n+1)
= J_{\parallel} + \lambda [u_j (n) - u_j (n+1)] .
$
Only the staggered part of the lattice displacement field along the chains, 
$u_j (n)$, couples to the spin-dimerization operator $\epsilon_j (n)$. 
We assume that 
$\omega_0 \gg J_{\perp}$, where $\omega_0$ is the phonon frequency at
$2k_F$, as is the case for the most ladder systems known.

In this limit, 
the $2k_F$-phonons can be treated in terms of a quantum Gaussian random field 
whose main effect is mediation of an
instantaneous effective coupling between the spin-dimerization fields
of each chain. We replace  the staggered parts of $u_j (n)$ 
by real scalar fields $\Delta_j (x)$,
$ u_j (n) = (-1)^n (a / 2 \lambda) \Delta_j (x)$, and  ignore  
 the kinetic energy
of vibrations. Then integrating over the displacement fields $\Delta_j$
we get the effective interaction 
\begin{eqnarray}
\Delta S_{\em eff} [ \epsilon] = 
- \frac{g^2 _0 (1 + \gamma/4)}{2 + \gamma} \int dx d \tau(\epsilon^2 _1 + \epsilon^2 _2)
 - \frac{g^2 _0 \gamma}{2 (2 + \gamma)} \int dx d \tau~ \epsilon_1 \epsilon_2 .
\label{eff.coupling}
\end{eqnarray} 
where $g_0 = \lambda / \sqrt{K_{\parallel} a}$ is the spin-phonon
coupling 
constant, $\gamma = K_{\perp} / K_{\parallel}$ is the ratio of 
the longitudinal and transverse
spring constants. The interchain coupling term  $\sim \epsilon_1 \epsilon_2$
in (\ref{eff.coupling}) is a relevant perturbation with dimension 1,
as opposed to terms $\sim \epsilon^2 _j$  which are only marginal and
therefore can be neglected\footnote{
Since the dimensionless spin-phonon coupling constant $g^2 _0 /v_s$ is
assumed to be small,
these terms only slightly renormalize the amplitude of the 
already existing marginally irrelevant
perturbation, $g {\bf J}_R \cdot {\bf J}_L~(g \sim 1)$, 
to the Gaussian fixed point
of the critical S = 1/2 spin chain  \cite{Aff-Hald}.}.
This explains the origin of the extra $V$-term
in the extended spin ladder model (\ref{gen.model}) 
and fixes the value of the constant
\begin{equation}
V \sim  - g^2 _0 \gamma / (2 + \gamma) < 0. \label{V}
\end{equation}

Let us consider various  phases of the model (\ref{gen.model}). 
We assume  that $V < 0$, while the sign of $J_{\perp}$
can be arbitrary. If  $J_{\perp} < 0$, the mass  $m_s > 0$,  
while the sign of $m_t$ depends on the strength of $|V|$ such that 
for small  $|V|$  
the signs of the two masses are opposite. All results obtained
in Ref.4 are applicable to this case, and we present 
them briefly for completeness.
The 
total (${\bf n}^{+} = {\bf n}_1 + {\bf n}_2$) and relative (${\bf n}^{-} = {\bf n}_1 - {\bf n}_2$)
staggered magnetization are given by\cite{Shelton1}:
\begin{eqnarray}
{\bf n}^+ &\sim& ( \sigma_1 \mu_2 \sigma_3 \sigma_0,~ \mu_1 \sigma_2 \sigma_3 \sigma_0,~ 
\sigma_1 \sigma_2 \mu_3 \sigma_0), \label{n+} \\
{\bf n}^- &\sim& ( \mu_1 \sigma_2 \mu_3 \mu_0, ~\sigma_1 \mu_2 \mu_3 \mu_0, ~\mu_1 \mu_2 \sigma_3
\mu_0 ), \label{n-}
\end{eqnarray} 
where $\sigma_a$ and  $\mu_a ~(a = 1,2,3)$ are order and disorder parameters of three, SU(2) degenerate
noncritical Ising models corresponding to the massive triplet of the Majorana fields $\xi_a$,
while $\sigma_0, ~\mu_0$ refer to the fourth, "singlet" Ising model ($\xi_0$). The Ising 
representation for
total and relative dimerization fields, $\epsilon_{\pm} = \epsilon_1 \pm \epsilon_2$, can be similarly
found to be
\begin{equation}
\epsilon_+ \sim \mu_1 \mu_2 \mu_3 \sigma_0, ~~~ \label{e+}
\epsilon_- \sim \sigma_1 \sigma_2 \sigma_3 \mu_0. \label{e-}
\end{equation}

The triplet mass determines the deviation of the 
Ising models from criticality:  
$m_t \sim (T - T_c)/T_c < 0$; this
corresponds to the ordered Ising phase with $\la  \mu_j \ra = 0,$
$\la \sigma_j \ra \neq 0 ~(j = 0,1,2,3)$, in which the two-point correlation functions are
given by \cite{Wu}
\begin{eqnarray}
\la \sigma ({\bf r}) \sigma ({\bf 0}) \ra &=& G_{\sigma} (m r) \simeq 
A \left[ 1 + \frac{1}{8 \pi (m r)^2} e^{- 2 |m| r} + O ( e^{- 4 |m| r}) \right] 
\label{Ising-cor1} \\
\la \mu({\bf r}) \mu ({\bf 0}) \ra &=&
G_{\mu} (m r)  \simeq \frac{A}{\pi} K_0 (|m| r) + O ( e^{- 3 |m| r}) , \label{Ising-cor2}
\end{eqnarray}
where $A$ is a nonuniversal parameter, $K_0 (x)$ is the MacDonald function,
and ${\bf r} = (x, v_s \tau)$.

In the region of parameters where $|m_t| \ll |m_s|$, the lowest (triplet) part of the spin ladder
spectrum describes universal properties of the S = 1 spin chain with
the conventional and biquadratic exchange, 
$
H = J \sum_n [ {\bf S}_n \cdot {\bf S}_{n+1} - \beta ({\bf S}_n \cdot {\bf S}_{n+1})^2 ],
$
near the critical (exactly integrable) point $\beta = 1$\cite{Babu}.
This correspondence in valid for any sign of $m_t$. The present case $m_t < 0$ describes the Haldane massive
phase ($\beta < 1$), and the magnitude of the Haldane gap
is given by $|m_t| \sim 1 - \beta$ \cite{Shelton1}.
All results obtained
in Ref.4 are applicable here:  
the leading asymptotics for the
spin correlation functions obtained from (\ref{n+}) -- (\ref{Ising-cor2})
\begin{equation}
\la {\bf n}^+ ({\bf r}) {\bf n}^+ ({\bf 0}) \ra 
\sim
K_0 (|m_t| r) [ 1 + O (e^{- 2 |m_t| r}) ],~~ 
\la {\bf n}^- ({\bf r}) {\bf n}^- ({\bf 0}) \ra 
\sim
\frac{1}{r^{3/2}} e^{- (2|m_t| + m_s) r}  \label{-asym} 
\end{equation}
reveal the role of ${\bf n}^+$ which determines the staggered magnetization of the effective
S = 1 chain. Since $K_0 (|m|r)$ is the real-space propagator of a free massive boson,
$\la {\bf n}^+ ({\bf r}) {\bf n}^+ ({\bf 0}) \ra$
contributes a $\delta$-peak to the imaginary
part of the dynamical spin susceptibility
corresponding to a massive triplet magnon.
At $q \sim 0$, $\chi'' (\omega, q)$ is determined by the correlations of slow components
of the total (${\bf J}_+ = {\bf J}_1 + {\bf J}_2 = - (i/2) (\vec{\xi} \times \vec{\xi})$) 
and relative (${\bf J}_- = {\bf J}_1 - {\bf J}_2 = i \vec{\xi} \xi_0$) magnetization, giving rise 
to thresholds at $2 |m_t|$ and $|m_t| + m_s$, respectively.
Similarly, one finds:
\begin{equation}
\la  \epsilon_- ({\bf r}) \epsilon_- ({\bf 0}) \ra 
\sim
K_0 (|m_s| r) \left[ 1 + O (e^{- 2 m_t r})  \right],~
\la  \epsilon_+ ({\bf r}) \epsilon_+ ({\bf 0}) \ra 
\sim
\frac{1}{r^{3/2}} e^{- 3 m_t r}, 
\end{equation}
implying that the singlet magnon is also a coherently propagating particle, --  an elementary
excitation of the relative dimerization. Excitations of the total dimerization have a 
$3 |m_t|$ threshold.
On increasing $|V|$, $|m_t|$ decreases while $m_s$ increases, the inequality $|m_t| < m_s$ thus
becoming stronger which makes the low-energy effective picture of the gapful Haldane phase of
the S = 1 spin chain only better. At $m_t = 0, ~m_s = 4|J_{\perp}|$ 
the triplet 
of the Majorana fields becomes massless, while 
the singlet Majorana fermion remains 
massive.
At this point
the correlation functions of the relative staggered magnetization and 
dimerization field
follow power laws:
$\la {\bf n}^- ({\bf r}) {\bf n}^- ({\bf 0}) \sim 
\la \epsilon_- ({\bf r}) \epsilon_- ({\bf 0}) \ra \sim  r^{- 3/4}.$ 
This critical point
belongs
to the universality class of  
the level k = 2 SU(2)-symmetric Wess-Zumino-Novikov-Witten model
with the central charge $c = 3/2$ \cite{Babu}.

Further increase of $|V|$ makes $m_t$ positive.
The change of sign of $m_tm_s$ amounts to the duality
transformation in the singlet ($\xi_0$) 
Ising system, implying that 
in formulas 
(\ref{n+}) -- (\ref{e-})
the order ($\sigma_0$) and disorder ($\mu_0$) parameters must be interchanged.
Moreover, since we are now in the disordered Ising phase 
($m_t \sim T - T_c > 0$), $<\mu_0> \neq 0, <\sigma_0> = 0$. 
the right-hand sides
of formulas (\ref{Ising-cor1}) and (\ref{Ising-cor2}) 
should also be interchanged.
As a result, the spin and dimerization 
correlation functions are now given by different expressions:
\begin{eqnarray}
\la {\bf n}^+ ({\bf r})\cdot {\bf n}^+ ({\bf 0})\ra  \sim K^2 _0 (m_t r), ~~
\la {\bf n}^- ({\bf r})\cdot {\bf n}^- ({\bf 0})\ra \sim K_0 (m_t r) K_0 (m_s r)
\label{n-new} \\
\la \epsilon_+ ({\bf r})  \epsilon_+ ({\bf 0}) \ra \sim  
C \left[ 1 + O (\frac{e^{- 2 m_{t,s}}}{r^2})  \right], ~~
\la \epsilon_- ({\bf r})  \epsilon_- ({\bf 0}) \ra \sim K^3 _0 (m_t r) K_0 (m_s r), 
\label{e-new}
\end{eqnarray}
where $C$ is a constant.

>From (\ref{e-new}) we conclude that the new phase is characterized by 
long-range dimerization ordering along each chain, with zero relative phase: 
$
\la \epsilon_1 \ra =  \la \epsilon_2\ra  = (1/2) \la \epsilon_+ \ra .
$
In the decoupling limit, $J_{\perp} = V = 0$, each Heisenberg chain  possesses a $Z_2$-symmetry:
this is the symmetry with respect to the interchange of  even and odd sublattices generated by 
one lattice spacing translation. The interchain coupling lowers this 
$Z_2 \times Z_2 $ down to $Z_2$, the symmetry
under {\it simultaneous} translations by $a_0$ on the both chains.
Thus, the onset of dimerization 
is associated with spontaneous breakdown of the residual
translational $Z_2$-symmetry, taking place when the interaction $|V|$ exceeds 
a critical value $|V_c| \sim J_{\perp}$. This phase coincides with the
dimerized phase of the generalized 
S = 1 spin chain with a biquadratic exchange\cite{Affleck1}.

In the dimerized phase the spin correlations undergo dramatic changes. 
Using long-distance asymptotics of the MacDonald functions, from 
(\ref{n-new}) 
we obtain:
\begin{eqnarray}
\Im m~ \chi_+ (\omega, \pi - q) 
&\sim& \frac{\theta (\omega^2 - q^2 - 4 m^2 _t)}{m_t 
\sqrt{\omega^2 - q^2 - 4 m^2 _t}}, \nonumber\\
\Im m~ \chi_- (\omega, \pi - q) &\sim&
\frac{\theta [\omega^2 - q^2 - (m_t + m_s)^2]}
{\sqrt{m_t m_s} \sqrt{\omega^2 - q^2 - (m_t + m_s)^2}}, \label{branch-cuts}
\end{eqnarray}
where $\pm$-signs refer to the case where the wave vector in the
direction perpendicular to the chains is equal to 0 and $\pi$
respectively.  We observe the disappearance of coherent
magnon poles in the dimerized  spin fluid; 
instead we find two-magnon thresholds at $\omega = 2 m_t$ 
and $\omega = m_t + m_s$, 
similar to the structure of $\chi'' (\omega, q)$ at small wave vectors in the Haldane fluid phase.
The fact that two massive magnons, each
with momentum $q \sim \pi$, combine to form a two-particle threshold, 
still at $q \sim \pi$ rather than $2\pi
\equiv 0$ is related to the fact that, in the dimerized phase with $2 a_0$-periodicity, the new Umklapp
is just $\pi$. 

To get a better understanding of the fact that in the dimerized phase the
spectral weight of the spin excitations 
is entirely incoherent, it is instructive to consider the limiting case $J_{\perp} = 0$. 
Since $\epsilon_{j}~(j = 1,2)$ are scalars in spin space, 
the model (\ref{gen.model}) displays
the $SU(2) \times SU(2) \approx SO(4)$ symmetry with respect 
to independent spin rotations
on each chain. Thus, the spectrum is described by O(4) quadruplet of massive Majorana fermions (with
the mass $m = - |V| (\pi a_0) |a_c|^2$), or equivalently, two noninteracting massive Dirac fermions, --
quantum solitons of two decoupled sine-Gordon models with the coupling constant $\beta^2 = 4 \pi$:
\begin{equation}  
H = H_+ + H_- = \sum_{s = \pm} \left[  \frac{v_s}{2} \left( \Pi^2 _s + (\p_x \Phi_s)^2 \right)
- \frac{m}{\pi a_0} \cos \sqrt{4 \pi} \Phi_s   \right]. \label{SG}
\end{equation} 
The massive Dirac fermions are in fact quantum
domain walls (kinks) connecting two $Z_2$-degenerate dimerized vacua of the two-chain system with
$\la \epsilon_+  \ra = \pm  |\epsilon_0|$.
This can be shown by the following simple consideration. A dimerization kink assumes
local changes $\epsilon_1 \rightarrow - \epsilon_1$, $\epsilon_2 \rightarrow - \epsilon_2$. This is
equivalent to simultaneous translations by $a_0$ on each chain, under which
$\Phi_j \rightarrow \Phi_j \pm \sqrt{\pi / 2}$. This, in turn, reduces to one of two possibilities:
$\Phi_+ \rightarrow \Phi_+ + \sqrt{\pi},$ $\Phi_- \rightarrow \Phi_-$, or
$\Phi_+ \rightarrow \Phi_+ ,$  $\Phi_- \rightarrow \Phi_- + \sqrt{\pi}.$
But $\sqrt{\pi}$ is just the period of the cosine potentials in (\ref{SG}). Therefore, a single kink
of the relative dimerization is nothing but a quantum sine-Gordon soliton (Dirac fermion) 
carrying a unit
topological charge, either in the $(+)$ or $(-)$ channel.
Now, in all physical excitations defined in the sector
with zero total topological charge, the solitons can appear only in pairs. Since two or more 
massive solitons
(Dirac fermions) cannot propagate coherently, and since there are no soliton-antisoliton bound 
states at
$\beta^2 = 4 \pi$ (free fermions!), one has to conclude that there will be no particle-like 
$\delta$-function peaks
in the spectral function of {\it any} local physical quantity of the system.

   Now let us consider the case $J_{\perp} > 0$, when $m_t > 0$
while $m_s$ may change its sign. If $m_s < 0$, the definitions (\ref{n+})-(\ref{e-}) still hold, 
but since $m_t \sim T - T_c > 0$, we are in the disordered Ising
phase. 
This case can be mapped onto
the $J_{\perp} < 0$ one by interchanging all order and disorder 
parameters which is equivalent to
the replacements ${\bf n}^+ \leftrightarrow {\bf n}^-$, $ \epsilon^+ \leftrightarrow \epsilon^-$.
As long as $|m_s| > m_t$, the triplet branch of 
the spectrum describes the Haldane phase
with the relative magnetization ${\bf n}^-$, forming effectively the staggered
component of the spin density in the S = 1 spin chain:
$\la {\bf n}^- ({\bf r}) \cdot {\bf n}^- ({\bf 0}) \ra 
\sim K_0 (|m_t| r).$

On increasing $|V|~|m_s|$ decreases, and the inequality $|m_s| > m_t$ will eventually be replaced
by the opposite one, $|m_s| < m_t$. One might argue that
as long as $|x| >> \xi_s$ (the maximal
correlation length  $\xi_s \sim v_s / |m_s|$), the asymptotic $r-$dependence 
of the ${\bf n}^-$-correlator
remains intact and the coherent magnon peak, though with a reduced amplitude, 
should still exist.
However, this is not so because the large-distance ($|{\bf r}| \gg
\xi_s$) 
asymptotics of the
${\bf n}^-$-correlator determines its Fourier transform 
$\la {\bf n}^- \cdot {\bf n}^-\ra_{q, \omega}$
at $|\pi - q|,|\omega| \ll m_s$, so that energies $|\omega| \sim m_t$ 
are not accessible.
At these energies
$\chi''(q \omega)$, is  mainly contributed 
by the asymptotics of the correlators
$\la {\bf n}^{\pm} ({\bf r}) \cdot {\bf n}^{\pm} ({\bf 0})  \ra$ in 
the range $\xi_t \geq |x| \gg \xi_s~(\xi_t \sim v_s / m_t)$, 
where $\mu_0$ cannot be
replaced by a constant:
\begin{equation}
\la {\bf n}^- ({\bf r}) {\bf n}^- ({\bf 0})  \ra \sim \frac{1}{r^{1/4}} K_0 (m_t r),~~
\la {\bf n}^+ ({\bf r}) {\bf n}^+ ({\bf 0})  \ra \sim \frac{1}{r^{1/4}} K^2 _0 (m_t r).
\label{new-asymp}
\end{equation}
Thus, even before reaching the critical point $m_s = 0$, 
the $\delta$-peak in dynamical susceptibility disappears, and
the effective picture of the
Haldane liquid breaks down when, due to softening of the
singlet mode, $|m_s|$ becomes comparable with $m_t$. 

At $m_s = 0~(|V| = 3 J_{\perp})$ the singlet excitations become gapless, and the intermediate asymptotics
(\ref{new-asymp}) are now exact. The dynamical spin susceptibility $\chi''_{\mp}(q \omega)$
displays thresholds at $\omega = m_t$ and $\omega = 2 m_t$, respectively. Near the thresholds
\begin{eqnarray}
\Im m~ \chi_+ (\omega, \pi - q) 
\sim \frac{\theta (\omega^2 - q^2 - 4 m^2 _t)}
{(\omega^2 - q^2 - 4 m^2 _t)^{3/8}},~~ 
\Im m~ \chi_- (\omega, \pi - q) \sim
\frac{\theta (\omega^2 - q^2 - m_t ^2)}
{ (\omega^2 - q^2 - m^2 _t )^{1/8}}, \label{branch-cuts1}
\end{eqnarray}
The correlation function of the total dimerization follows a power law $
\la \epsilon^+ ({\bf r}) \epsilon^+ ({\bf 0}) \ra \sim r^{-1/4}$. 
corresponding to the Ising order parameter at criticality. 
Thus the critical point $m_s = 0$
belongs to the Ising universality class (with central charge $ c =
1/2$) 
and signals a transition
to a spontaneously dimerized phase  with $\la \epsilon^+ \ra \neq 0$. In the dimerized phase
($m_s > 0$) the singlet mass gap is always the smallest indicating
that 
the spin ladder is not in the regime
of an effective S = 1 spin chain.

The Haldane liquid has   a nonlocal topological (string) order
parameter\cite{denNijs} whose nonzero value is associated with the breakdown 
of a hidden discrete ($Z_2 \times Z_2$) symmetry. In the two-chain realization
of the Haldane S=1 phase the string order parameter is expressed in
terms of the
Ising order and disorder variables\cite{Shelton1}:
\begin{equation}  
\lim_{|x - y| \rightarrow \infty} \la O_{\em string} (x,y) \ra \sim 
\la \sigma_1 \ra^2 \la \sigma_2 \ra^2 + \la \mu_1 \ra^2 \la \mu_2 \ra^2 . \label{string}
\end{equation}
(notice that the singlet mode does not appear in this expression). 
Since the Ising systems are noncritical in both phases, it follows from
(\ref{string}) that the string order parameter
will be nonzero in the dimerized phases as well, vanishing only 
at the critical points $m_t = 0$
and $m_s = 0$.

Thus we have demonstrated that the Haldane spin liquid is not the
only possible phase of a disordered magnet. 
The distinctive features of another
- dimerized phase,
which can be tested by  inelastic neutron scattering and NMR experiments is the
absence of coherent single-magnon modes and the $\pi$-periodicity of the 
spin excitation spectrum, as opposed to the undimerized phase where the ratio
of the energy gaps at $q \sim 0$ and $q \sim \pi$ is 2.
The Haldane- and dimerized phases are separated by critical points, either
of the Ising type or belonging to the universality class of the critical S = 1
quantum spin chain.

We are grateful to Yu Lu, Ilya Krive and Alexander Gogolin 
for valuable discussions.
One of us (A.A.N.) acknowledges hospitality and support from 
the International Centre for
Theoretical Physics.


\end{document}